\documentclass[pra,aps,amsmath,amssymb,twocolumn,showpacs]{revtex4}
\usepackage{graphicx}
\usepackage{color}
\usepackage{amsmath}
\usepackage{amssymb}
\newcounter{abcd}
\newcommand{\be}{\begin{equation}}
\newcommand{\ee}{\end{equation}}
\newcommand{\bd}{\begin{displaystyle}}
\newcommand{\ed}{\end{displaystyle}}
\newcommand{\ba}{\begin{eqnarray}}
\newcommand{\ea}{\end{eqnarray}}
\DeclareMathSymbol{\lang}{\mathord}{symbols}{"68}
\DeclareMathSymbol{\rang}{\mathord}{symbols}{"69}
\DeclareMathSymbol{\openbra}{\mathord}{symbols}{"68}
\DeclareMathSymbol{\closeket}{\mathord}{symbols}{"69}
\DeclareMathOperator{\Rre}{Re}
\DeclareMathOperator{\Iim}{Im}

\newcommand{\ket}[1]{{| #1 \closeket}}

\newcommand{\aver}[1]{{\lang #1 \rang}}
\begin{document}
\title{Theory of chaotic atomic transport in an optical lattice}
\author{V.Yu. Argonov and S.V. Prants}
\affiliation{Laboratory of Nonlinear Dynamical Systems,\\
V.I.Il'ichev Pacific Oceanological Institute\\
of the Russian Academy of Sciences, 43 Baltiiskaya st.,\\
690041 Vladivostok, Russia}

\begin{abstract}
A semiclassical theory of chaotic atomic transport in a one-dimensional 
nondissipative optical lattice is developed. Using the basic equations of motion 
for the Bloch and translational atomic variables, we derive a stochastic map 
for the synchronized component of the atomic dipole moment that determines the center-of-mass 
motion. We find the analytical relations between the atomic and lattice 
parameters under which atoms typically alternate between flying through 
the lattice and being trapped in the wells of the optical potential. 
We use the stochastic map to derive formulas for the probability 
density functions (PDFs) for the flight and trapping events. Statistical 
properties of chaotic atomic transport strongly depend on the relations between 
the atomic and lattice parameters. We show that there is a good quantitative 
agreement between the analytical 
PDFs and those computed with the stochastic map and the basic equations of 
motion for different ranges of the parameters. Typical flight and 
trapping PDFs are shown to be broad distributions with power law ``heads'' 
with the slope $-1.5$ and exponential ``tails''. The lengths of the 
power law and exponential parts of the PDFs depend on the values of the 
parameters and can be varied continuously. 
We find analytical 
conditions, under which deterministic atomic transport has fractal properties, 
and explain a hierarchical structure of the dynamical fractals.  
\end{abstract}
\pacs{42.50.Vk, 05.45.Mt, 05.45.Xt}
\maketitle

\section{Introduction}
The transport properties of cold atoms in optical
lattices depend on the lattice and atomic parameters
and can be very diverse. The optical lattice is a periodic 
structure of micron-sized potential wells which is created
by a laser standing wave made of two counterpropagating
laser beams. Dilute atomic samples with negligibly
small atom-atom interactions are used to probe
experimentally single-atom phenomena.
The atomic motion can take form of ballistic transport,
oscillations in wells of the optical potential, Brownian
motion, random walks, L\'evy flights, and chaotic
transport.

Most of the experimental \cite{BB02,BB94,JD96,KS97,KO98,SL99,GR01} 
and theoretical \cite{BB02,GR01,ME96,GP97,SS99,L04} works on
the atomic transport in optical lattices have been
done in the context of laser cooling of atoms. One particularly interesting 
aspect of these works is the discovery of anomalous transport properties 
of cold atoms and L\'evy flights \cite{BG90,SZ93}. A L\'evy flight is a 
random process in the space (time) domain with the distribution of the length 
(duration) of flight events that is given by a L\'evy law possessing an 
algebraically decaying ``tail'' and infinite variance. As a consequence, 
the atomic trajectories (time series) have self-similar (fractal) nature 
and the probability to find superlong flights is not negligibly small as 
in a case of normal diffusion. 

In the context of sub-recoil laser cooling, the L\'evy flights have 
been found experimentally \cite{BB02,BB94} in the distributions 
of trapping and escape times 
for ultracold atoms trapped in a momentum state close to the dark state. 
It was shown in Refs. \cite{BB02,BB94} that not only the variance but the mean 
time for atoms to leave the trap is infinite. 

The anomalous atomic transport of cold atoms in optical lattices is of a different nature. 
A cold atom in an optical lattice can be trapped in the potential wells and 
it can move over many wavelengths in dependence on whether its energy is 
below or above the potential barrier. Transitions between these events are 
stochastically induced by the cooling force and its fluctuations 
(heating due to randomness of spontaneous emission and 
fluctuations of the atomic dipole moment). Measurement of the trajectory of 
a single cold ion in a one-dimensional optical lattice, by tracing its position through 
fluorescent photons, have demonstrated L\'evy flights \cite{KS97}. By decreasing 
the optical potential depth, a change of the transport characteristics 
from diffusive to quasiballistic was observed \cite{KS97}.  

The anomalous properties of atomic transport discussed briefly above are 
associated with random atomic recoils in spontaneous emission 
processes which are generally inevitable in any cooling scheme and which 
make the transport looks like a random walk. 
However, the problem may be considered not in a laser cooling but 
in a more general context as a study of deterministic motion of atoms with comparatively 
large momentum interacting with a standing-wave light field. It is a
fundamental nonlinear interaction between different
degrees of freedom (the translational and internal atomic
degrees of freedom and the field ones) that can be
treated in a Hamiltonian form when neglecting any
losses. In this context spontaneous emission may be considered as a 
noise imposed on the coherent atomic dynamics. 
It has been predicted in Refs. \cite{PK01,PS01}
that, besides the well-known transport properties of atoms
in optical lattices, there should exist a deterministic chaotic
transport with a complicated alternation of atomic
oscillations in wells of the optical potential and
atomic flights over many potential wells when the atom may change 
the direction of motion many times. This phenomenon looks like a
random walk but it should be stressed
that it may occur without any random fluctuations
of the lattice parameters and any noise like
spontaneous emission. The deterministic chaotic transport is a
result of chaotic atomic dynamics in the standing-wave
field that means an exponential sensitivity of the
internal and translational atomic variables to
small variations in the lattice parameters and/or
initial conditions. The chaotic transport and its
manifestations like atomic dynamical fractals may
occur both in classical \cite{Pr02,JETP,PEZ,G02} and quantized light
fields \cite{PU03,PU06}. Spontaneous emission
events interrupt the coherent atomic dynamics
in random instants of time and may give rise anomalous
statistical properties of atomic transport \cite{AP06}.

In our previous papers \cite{JETP,PEZ,JRLR,PU06}
we have found chaotic atomic transport and dynamical fractals
in numerical experiments and studied its properties and manifestations. 
The ranges
of the lattice and atomic parameters and initial conditions, 
for which the center-of-mass motion may be chaotic, have been established. 
In this paper we develop a
semiclassical Hamiltonian theory of the chaotic atomic transport in a
one-dimensional optical lattice and confirm the
analytical results by the numerical simulation.
In \setcounter{abcd}{2}Sec.~\Roman{abcd} we derive the basic equations 
of motion and give
the result of computation of the maximum Lyapunov
exponent whose positive values determine
the ranges of the atom-field detuning $\Delta$ and initial
atomic momentum $p_0$ for which chaotic transport
occurs. \setcounter{abcd}{3}Sec.~\Roman{abcd} briefly reviews distinct regimes of motion.
Using approximate solutions of the basic equations, we
construct in \setcounter{abcd}{4}Sec.~\Roman{abcd} a stochastic map 
for the synchronized component of the atomic dipole moment  $u$ 
that determines the chaotic transport. 
In \setcounter{abcd}{5}Sec.~\Roman{abcd}
we introduce a simple illustrative model of random walking
of the quantity $\arcsin u$ on a circle. Depending on
the relations between the lattice and atomic parameters, the transport
properties may be diverse. We
find in \setcounter{abcd}{5}Sec.~\Roman{abcd} the condition under which the probability
density functions (PDFs) for the flights and
trappings of the atoms either purely exponential or
have prominent power law slopes. We derive the PDFs
analytically and compare the results with simulation of the
stochastic map and the basic equations. The results obtained in 
\setcounter{abcd}{4}Sec.~\Roman{abcd}
are used in \setcounter{abcd}{6}Sec.~\Roman{abcd} to find
the conditions for appearing dynamical atomic fractals
and to explain their structure. Finally, 
\setcounter{abcd}{7}Sec.~\Roman{abcd}gives conclusions. 

\section{Hamilton-Schr\"odinger equations of motion}

We consider a two-level atom with mass $m_a$ and transition
frequency $\omega_a$, moving with the momentum $P$ along the axis $X$
in a one-dimensional classical standing laser wave with the frequency $\omega_f$
and the wave vector $k_f$. In the frame,
rotating with the frequency $\omega_f$, the Hamiltonian is
the following:
\begin{equation}
\hat H=\frac{P^2}{2m_a}+\frac{1}{2}\hbar(\omega_a-\omega_f)\hat\sigma_z-
\hbar \Omega\left(\hat\sigma_-+\hat\sigma_+\right)\cos{k_f X}.
\label{Jaynes-Cum}
\end{equation}
Here $\hat\sigma_{\pm, z}$ are the Pauli operators which describe the transitions
between lower, $\ket{1}$, and upper, $\ket{2}$, atomic states,
$\Omega$ is the Rabi frequency which is proportional to the square
root of the number of photons in the wave $\sqrt{n}$.
The laser wave is assumed to be strong enough ($n\gg 1$),
so we can treat the field classically.
The simple wavefunction for the electronic degree of freedom is
\begin{equation}
\ket{\Psi(t)}=a(t)\ket{2}+b(t)\ket{1},
\label{Psi}
\end{equation}
where $a$ and $b$ are the complex-valued probability amplitudes to find the
atom in the states $\ket{2}$ and $\ket{1}$, respectively.
Using the Hamiltonian (\ref{Jaynes-Cum}), we get the Schr\"odinger equation
\begin{equation}
\begin{array}{l}
\begin{displaystyle}
i\frac{da}{dt}=\frac{\omega_a-\omega_f}{2}a-\Omega b\cos k_fX,\
\end{displaystyle}
\\
\\
\begin{displaystyle}
i\frac{db}{dt}=\frac{\omega_f-\omega_a}{2}b-\Omega a\cos k_fX.\
\end{displaystyle}
\end{array}
\label{sysa}
\end{equation}
Let us introduce
instead of the complex-valued probability amplitudes $a$ and $b$ the
following real-valued variables:
\begin{equation}
\begin{gathered}
u\equiv 2\Rre\left(ab^*\right),\quad
v\equiv -2\Iim\left(ab^*\right),\\
z\equiv \left|a\right|^2-\left|b\right|^2,
\end{gathered}
\label{uvz_def}
\end{equation}
where $u$ and $v$ are a synchronized (with the laser field) and a quadrature 
components of the atomic electric dipole moment, respectively, and $z$ is 
the atomic population inversion.

In the process of emitting and
absorbing photons, atoms not only change their internal electronic states
but their external translational states change as well due to the photon
recoil. If the atomic mean momentum is large as compared to the photon momentum
$\hbar k_f$, one can describe the translational degree
of freedom classically. The position and momentum of a point-like atom
satisfy classical Hamilton equations of motion. Full dynamics in the
absence of any losses is now governed by the Hamilton-Schr\"odinger
equations for the real-valued atomic variables
\begin{equation}
\begin{aligned}
\dot x&=\omega_r p,
\\
\dot p&=- u\sin x,
\\
\dot u&=\Delta v,
\\
\dot v&=-\Delta u+2 z\cos x,
\\
\dot z&=-2 v\cos x,
\end{aligned}
\label{mainsys}
\end{equation}
where $x\equiv k_f X$ and $p\equiv P/\hbar k_f$ are classical
atomic center-of-mass position and momentum, respectively.
Dot denotes differentiation with respect to the dimensionless time $\tau\equiv \Omega t$.
The normalized recoil frequency, $\omega_r\equiv\hbar k_f^2/m_a\Omega\ll 1$,
and the atom-field detuning,
$\Delta\equiv(\omega_f-\omega_a)/\Omega$, are the control parameters.
The system has two integrals of motion, namely the total energy
\begin{equation}
H\equiv\frac{\omega_r}{2}p^2-u\cos x-\frac{\Delta}{2}z,
\label{H}
\end{equation}
and the Bloch vector
\begin{equation}
u^2+v^2+z^2=1.
\label{R}
\end{equation}
The conservation of the Bloch
vector length immediately follows from
Eqs. (\ref{uvz_def}).

Equations of motion similar to the set (\ref{mainsys})
were obtained in our previous papers \cite{PK01,PS01,JETP} in order to describe
the interaction between a two-level atom and the cavity radiation field
in the strong-coupling limit. Taking into account a back reaction of the
atom on the radiation field and within the semiclassical approximation, we
were able to get the corresponding version of the Hamilton-Schr\"odinger
equations for the atomic position $x$, momentum $p$, population
inversion $z$, and two combined atom-field variables which were denoted
by the same letters $u$ and $v$ as the atomic dipole-moment 
components in Eqs. (\ref{mainsys}) Those equations has
been shown in Refs. \cite{PK01,PS01,JETP} to have a positive Lyapunov exponent
in a wide range of the control parameters and initial atomic
momentum $p_0$. It implies dynamical chaos in the usual sense
of exponential sensitivity to small changes in initial conditions and/or
control parameters. The same should be valid with the set of equations (\ref{mainsys})
describing the  different physical situation --- a two level atom
in open space with a strong standing-wave field.

\begin{figure}[htb]
\begin{center}
\includegraphics[width=0.45\textwidth,clip]{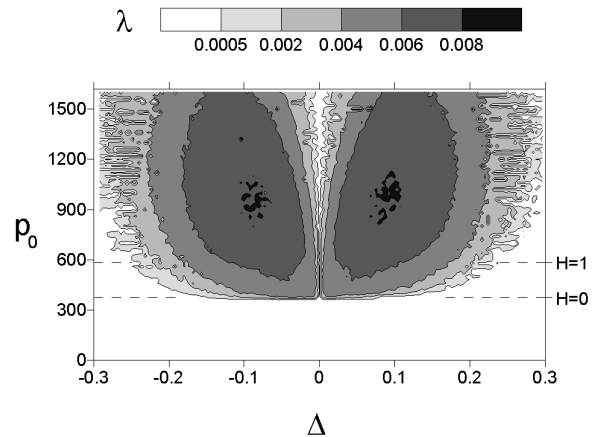}
\end{center}
\caption{Maximum Lyapunov exponent $\lambda$ vs atom-field detuning 
$\Delta$ (in units of the laser Rabi frequency $\Omega$)
and initial atomic momentum $p_0$ (in units of the photon momentum 
$\hbar k_f$): $\omega_r=10^{-5}$, $u_0=z_0=0.7071$, $v_0=0$.}
\label{fig1}
\end{figure}

Equations (\ref{mainsys}) constitute a nonlinear Hamiltonian
autonomous system with two and half degrees of freedom which, owing to two integrals of motion, move on a three-dimensional
hypersurface with a given energy value $H$. In general, motion in a three-dimensional phase space in characterized by a positive
Lyapunov exponent $\lambda$, a negative exponent equal in magnitude to the positive one,
and zero exponent. The sum of all Lyapunov exponents of a Hamiltonian system is
zero \cite{LL}. The maximum Lyapunov exponent characterizes the mean rate of the
exponential divergence of initially close trajectories,
\begin{equation}
\lambda\equiv\lim\limits_{\tau\to\infty}\lambda(\tau),\quad
\lambda(\tau)\equiv\lim\limits_{d(0)\to 0}\frac{1}{\tau}\ln\frac{d(\tau)}{d(0)},
\end{equation}
and serves as a quantitative measure of dynamical chaos in the system.
Here, $d(\tau)$ is a distance (in the Euclidean sense) at time
$\tau$ between two trajectories close to each other at initial time moment
$\tau=0$.
The result of computation
of the maximum Lyapunov exponent in dependence on
the detuning $\Delta$ and the initial atomic momentum $p_0$ is shown in
Fig.~\ref{fig1}. Color in the plot marks the value of the maximum Lyapunov exponent $\lambda$.
In white regions the values of $\lambda$ are almost zero, 
and the atomic motion is regular in the corresponding ranges 
of $\Delta$ and $p_0$. In
shadowed regions positive values of $\lambda$ imply unstable motion.

In all numerical simulations we shall use the normalized
value of the recoil frequency equal to $\omega_r=10^{-5}$. The initial atomic
position is taken to be $ x_0=0$. The detuning $\Delta$ will be varied
in a wide range, and the Bloch variables are restricted by the length
of the Bloch vector (\ref{R}). It should be noted that  we use in this paper
the normalization to the laser Rabi frequency $\Omega$, not to the vacuum (or
single-photon) Rabi frequency as it has been done in our previous papers
\cite{PK01,PS01,Pr02,JETP}. So the ranges of the normalized control
parameters, taken in this paper, differ from those in the cited papers.
Figure~\ref{fig1} demonstrates that the center-of-mass motion becomes unstable 
if the dimensionless momentum exceeds the value $p_0 \approx 300$ that 
corresponds (with our normalization) to the atomic velocity $v_a\approx 1$ m/s 
of a cesium atom in the field with the wavelength close to the transition 
wavelength $\lambda_a\simeq 852$~nm.

\section{Regimes of motion}
\subsection{Regular atomic motion at exact atom-field resonance}

The case of exact resonance, $\Delta=0$, was considered in detail in Ref. \cite{PS01,JRLR}.
Now we briefly repeat the simple results for the sake of self-consistency. 
At zero detuning, the variable $u$ becomes a constant, $u=u_0$,
and the fast ($u$, $v$, $z$) and slow ($x$, $p$) variables are separated
allowing one to integrate exactly the reduced equations of motion.
The total energy is equal to
\be
H_0=\frac{\omega_r}{2}p^2-u\cos x,
\label{8}
\end{equation}
where $u=u_0$.
The center-of-mass atomic motion in this
spatially periodic potential of the standing wave
is described by the simple nonlinear equation for a free physical pendulum
\begin{equation}
\label{7}
\ddot x+\omega_r u_0\sin x=0,
\end{equation}
and does not depend on evolution of the internal degrees of freedom.

The translational motion is trivial when $u_0$ is zero. The atom
moves in one direction with a constant velocity, and
the Rabi oscillations are modulated by the standing wave.
Equations (\ref{8}) and (\ref{7}) describe
the atomic motion in the simple cosine potential $u_0 \cos x$
with three types of trajectories which are possible in dependence on 
the value of the energy
$H$: oscillator-like motion in a potential well if $H_0< u_0$
(atoms are trapped by the standing-wave field \cite{Letokhov}),
motion along the separatrix if $H_0= u_0$, and ballistic-like motion if $H_0> u_0$.
Exact solutions of Eq. (\ref{7}) are easily found in terms of elliptic
functions (see \cite{PS01,JRLR}).

As to internal atomic evolution, it depends on the translational degree of freedom
since the strength of the atom-field coupling depends on the position of
atom in a periodic standing wave.
At $\Delta=0$, it is easy to find the exact solutions of Eqs. (\ref{mainsys})
\begin{equation}
\begin{aligned}
v(\tau)=\pm\sqrt{1-u^2}\ \cos\left(2\int\limits_0^\tau \cos  x d\tau'+
\chi_0\right),\\
z(\tau)=\mp\sqrt{1-u^2}\ \sin\left(2\int\limits_0^\tau \cos  x d\tau'+
\chi_0\right),
\label{vz}
\end{aligned}
\end{equation}
where $u=u_0$, and $\cos[ x(\tau)]$ is a given function of
the translational variables only which can be found with the help of
the exact solution for $x$ \cite{PS01,JRLR}. The sign of $v$ is equal to that for the initial
value $z_0$ and
\be
\chi_0\equiv\mp\arcsin\frac{z_0}{\sqrt{1-u_0^2}}
\end{equation}
is an integration constant. The internal energy
of the atom, $z$, and its quadrature dipole-moment component $v$ 
could be considered as frequency-modulated signals 
with the instant frequency $2\cos[ x(\tau)]$ and the modulation frequency
$\omega_r p(\tau)$,
but it is correct only if the maximum value of the first frequency is much greater than the value of the second one,
i. e., for $|\omega_r p_0|\ll 2$.

\subsection{Chaotic atomic transport off the resonance}

\begin{figure}[htb]
\begin{center}
\includegraphics[width=0.45\textwidth,clip]{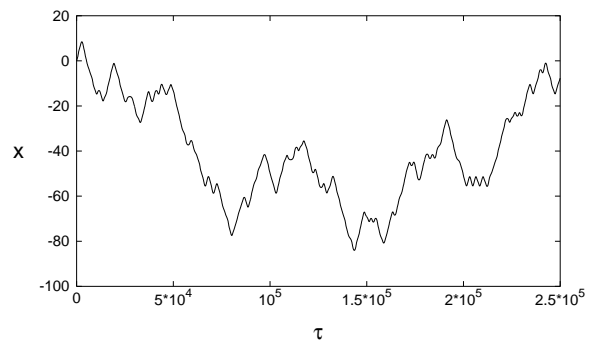}
\end{center}
\caption{Typical atomic trajectory in the regime of chaotic
transport: $x_0=0$, $p_0=300$, $z_0=-1$, $u_0=v_0=0$, $\omega_r=10^{-5}$,
$\Delta=-0.05$. Atomic position is shown in units of $k_f^{-1}$, time $\tau$
in units of $\Omega^{-1}$.}
\label{fig2}
\end{figure}

In Fig.~\ref{fig1} we show the $\lambda$-map
in the space of the initial momentum $p_0$ and detuning $\Delta$ values.
The maximum Lyapunov exponent $\lambda$ depends both on
the parameters $\omega_r$ and $\Delta$, and on initial conditions
of the system (\ref{mainsys}). It is naturally to expect
that off the  resonance atoms with comparatively small values
of the initial momentum $p_0$ will be at once trapped in the first well
of the optical potential, whereas those with large values of $p_0$ will
fly through. The question is what will happen with
atoms, if their initial kinetic energy will be close to the maximum of the optical potential.
Numerical experiments demonstrate that such atoms will wander in the
optical lattice with alternating trappings in the wells of the optical
potential and flights over its hills. The direction of the center-of-mass
motion of wandering atoms may change in a chaotic way (in the sense of exponential
sensitivity to small variations in initial conditions).
A typical
chaotic atomic trajectory is shown in Fig.~\ref{fig2}.

It follows from (\ref{mainsys}) that the translational motion of the atom
at $\Delta \neq 0$ is described by the equation of a nonlinear physical
pendulum with the frequency modulation
\begin{equation}
\ddot x+\omega_r  u(\tau)\sin x=0,
\label{12}
\end{equation}
where $u$ is a function of all the other dynamical variables.

\section{Stochastic map for chaotic atomic transport}

Chaotic atomic transport occurs even if the normalized
detuning is very small, $|\Delta|\ll 1$ (Fig.~\ref{fig1}). 
Under this condition, we will derive in this section
approximate equations for the center-of-mass motion. The atomic 
energy at $|\Delta|\ll 1$ is given with a good accuracy by the simple
resonant expression (\ref{8}). Returning to the basic set of the equations of motion 
(\ref{mainsys}),
we may neglect the first term in the fourth equation since it is very small
as compared with the second one there. However, we cannot now exclude
the third equation from the consideration. Using the solution
(\ref{vz}) for $v$, we can transform this equation as
\begin{equation}
\dot u=\pm\Delta\sqrt{1-u^2}\ \cos\chi,
\label{u}
\end{equation}
where
\begin{equation}
\chi\equiv 2\int\limits_0^\tau \cos  x d\tau'+ \chi_0.
\end{equation}
Far from the nodes of the standing wave, Eq. (\ref{u}) can be approximately 
integrated under the additional condition, 
$|\omega_r p|\ll 1$, which is valid for the ranges of the parameters and the
initial atomic momentum where chaotic transport occurs. Assuming $\cos x$
to be a slowly-varying function in comparison with the
function $\cos\chi$, we obtain far from the nodes the approximate solution
for the $u$-component of the atomic dipole moment 
\begin{equation}
u\approx\sin\left(\pm\frac{\Delta}{2\cos x} \sin \chi+C\right),
\end{equation}
where $C$ is an integration constant. Therefore, the amplitude
of oscillations of the quantity $u$ for comparatively slow
atoms ($|\omega_r p|\ll 1$) is small and of the order of $|\Delta|$
far from the nodes. 

\begin{figure}[htb]
\begin{center}
\includegraphics[width=0.45\textwidth,clip]{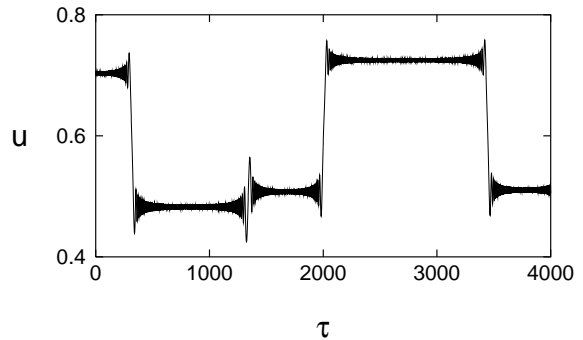}
\end{center}
\caption{Typical evolution of the atomic dipole-moment component 
$u$ for a comparatively slow and slightly detuned atom: 
$x_0=0$, $p_0=550$, $v_0=0$, $u_0=z_0=0.7071$,
$\omega_r=10^{-5}$, $\Delta=-0.01$.}
\label{utraj}
\end{figure}

It follows from the third and forth equations in the set (\ref{mainsys}) 
that $u$ satisfies to the equation of motion for a driven harmonic 
oscillator with the natural frequency $| \Delta |$ and the driving 
force $2z\cos x$ whose frequency is space and time dependent. 
At $| \Delta | =0$, the synchronized component of the atomic dipole moment 
$u=$ is a constant whereas the other Bloch variables  
$z$ and $v$ oscillate in accordance with the solution (\ref{vz}). 
At $|\Delta | \neq 0$ and far from the nodes, the variable $u$ performs 
shallow oscillations for the natural frequency $| \Delta |$ is small as 
compared with the Rabi frequency. However,    
the behavior of $u$ is expected to be very special when
an atom approaches to any node of the standing wave since near the node 
the oscillations of the atomic population inversion $z$ slow down and 
the corresponding driving frequency becomes close to the resonance with the 
natural frequency. As a result,
sudden ``jumps'' of the variable $u$ are expected to occur near the nodes.
This conjecture is supported by the numerical simulation. In  
Fig.~\ref{utraj} we show a typical behavior of the variable $u$ for a
comparatively slow and slightly detuned atom.
The plot clearly demonstrates sudden ``jumps'' of $u$ near
the nodes of the standing wave and small oscillations between the nodes.

Approximating the variable $u$ between the nodes by
constant values, we can construct a discrete mapping $u_m=f(u_{m-1})$,
where $u_m$ is a value of $u$ just after the $m$-th node crossing    
(including multiple crossings of the same node). In order to
estimate the values of $u$ just after the crossing, one
needs to integrate Eq. (\ref{u}) near a node. Since $u$ changes
largely in a small region near nodes (so small that the atomic momentum
has no time to change its value significantly) we may use
the Raman-Nath approximation of the constant velocity. In this
approximation, we have
\begin{equation}
\chi\simeq\frac{2}{\omega_r p_{\rm{node}}}\sin x+\chi_0.
\end{equation}
Substituting this expression into Eq. (\ref{u}), we can integrate
it in the interval $0\leqslant x\leqslant\pi$ (which comprises only one node)
and obtain the value of $u$ just after crossing the first node
\begin{equation}
\begin{aligned}
\ &u_1\approx\sin\left(\arcsin u_0\pm\frac{\Delta}{\omega_r p_{\rm{node}}}
\int\limits_{0}^{\pi}\cos\chi dx\right)=\\
\ &\ =\sin\left(\arcsin u_0\pm\frac{\Delta\pi}{\omega_r p_{\rm{node}}\sqrt{1-u_0^2}}\right.\times\\
\ &\ \times\left[v_0 J_0\left(\frac{2}{\omega_r p_{\rm{node}}}\right)
+\left.z_0 E_0\left(\frac{2}{\omega_r p_{\rm{node}}}\right)\right]\right),
\label{JE}
\end{aligned}
\end{equation}
where 
\begin{equation}
p_{\rm{node}}\equiv \sqrt\frac{2H}{\omega_r}
\end{equation}
is the value of the atomic momentum at the instant when the atom crosses 
a node (which is the same with a given value of the energy $H$  
for all the nodes),
$J_0$ and $E_0$ are zero-order Bessel and Weber functions, respectively.
In the limit of the large argument
$2/(\omega_r p_{\rm{node}})$, both the functions have a harmonic
asymptotics \cite{Emde}, and the expression (\ref{JE}) reduces to the form
\begin{equation}
\begin{aligned}
u_1\approx\sin\left(\pm\frac{\Delta}{\sqrt{1-u_0^2}}
\right.&\left[\sqrt\frac{\pi}{\omega_r p_{\rm{node}}}\right.\times\\
\times\left(v_0\cos\left(\frac{2}{\omega_r p_{\rm{node}}}-\frac{\pi}{4}\right)\right.-
\ &\left.z_0\sin\left(\frac{2}{\omega_r p_{\rm{node}}}-\frac{\pi}{4}\right)\right)-\\
\ &-\left.\left. z_0\right]+\arcsin u_0\right).
\end{aligned}
\label{u_1}
\end{equation}
It is the deterministic solution obtained with
the approximations mentioned above. It should be stressed that the solution contains
trigonometric functions with large values of the arguments which are inversely
proportional to the atomic momentum. In the Raman-Nath approximation,
we take $p\simeq p_{\rm{node}}$. In fact, even small deviations from
this mean value may result in large changes in the magnitude of the trigonometric
functions. Therefore, they can be treated as, practically random
variables in the range $[-1,1]$. Beyond the Raman-Nath approximation, 
the value of the atomic momentum $p$ depends
on the value of $u_m$ which changes every time when the atom crosses a node. 
So, we can replace arguments of the trigonometric functions by random
variables. Finally, we introduce the stochastic map
\begin{equation}
\begin{aligned}
\ &u_{m}\equiv\sin\left(\Delta\sqrt\frac{\pi}{\omega_r p_{\rm{node}}}
\sin\phi_{m}+\arcsin u_{m-1}\right)=\\
\ &=\sin\left(\Delta\sqrt\frac{\pi}{\omega_r p_{\rm{node}}}\sum\limits_{j=1}^{m}
\sin\phi_j+\arcsin u_0\right),
\end{aligned}
\label{u_m}
\end{equation}
where $\phi_m$ are random phases to be chosen in the range
$[0,2\pi]$. When deriving this map, we neglected the term $z_0$ in Eq. (\ref{u_1})
which is small as compared with the factor 
$\sqrt{\pi/\omega_r p_{\rm{node}}}$. 

With given values of $\Delta$,
$\omega_r$, and $p_{\rm{node}}$, the map (\ref{u_m}) has been shown
numerically to give a satisfactory probabilistic distribution of
magnitudes of changes in the variable $u$ just after crossing
the nodes. The stochastic map (\ref{u_m}) is valid under the assumptions
of small detunings ($|\Delta|\ll 1$) and comparatively slow atoms
($|\omega_r p|\ll 1$). Furthermore, it is valid only for those ranges
of the control parameters and initial conditions where the motion
of the basic system (\ref{mainsys}) is unstable. For example,
in those ranges where all the Lyapunov exponents are zero, $u$ becomes
a quasi-periodic function and cannot be approximated by the map.

The stochastic map (\ref{u_m}) allows to reduce the basic 
set of equations of motion (\ref{mainsys}) to the following
effective equations of motion:
\begin{equation}
\begin{aligned}
\ &\dot x&=\ &\omega_r p,
\\
\ &\dot p&=\ &- u_m\sin x,
\\
\ &\dot m&=\ &\omega_r p_{\rm{node}}\ |\delta(\cos x)|,
\end{aligned}
\label{redsys}
\end{equation}
where $u_m$ is found from Eq. (\ref{u_m}). The third equation
in the set (\ref{redsys}) gives a correspondence between
the continuous evolution of the atomic motion and
a discrete crossing number $m$. The
integration of the delta function $\delta(\cos x)$ over time at points with $\cos x=0$
gives $\pm 1/(\omega_r p_{\rm{node}})$ in dependence on the direction 
of motion and whether the serial number of the node is even or odd. 
Since we calculate absolute values of the delta function,
it is easy to show that $m$ is a constant if $\cos x\ne 0$
and increases by one if $\cos x=0$.

\section{Statistical properties of chaotic transport}
\subsection{Model for chaotic atomic transport}

With given values of the control parameters and the energy $H$,
the center-of-mass motion is determined by the values of $u_m$
(see Eq. (\ref{12})). One can obtain from the expression for the
energy (\ref{8}) the conditions under which atoms continue to
move in the same direction after crossing a node
or change the direction of motion not reaching the
nearest antinode. Moreover, as in the resonance case,
there exist atomic trajectories along which atoms 
move to antinodes with the velocity going
asymptotically to zero. It is a kind of separatrix-like motion with 
an infinite time of reaching the stationary points.

The conditions for different regimes of motion depend on
whether the crossing number $m$ is even or odd.
Motion in the same direction occurs at $(-1)^{m+1}u_m<H$, 
separatrix-like motion --- 
at $(-1)^{m+1}u_m=H$, and turns --- at $(-1)^{m+1}u_m>H$.
It is so because even values of $m$ correspond to $\cos x>0$,
whereas odd values --- to $\cos x<0$. The quantity $u$
during the motion changes its values in a random-like manner
(see Fig.~\ref{utraj}) 
taking the values which provide the atom either to prolong the motion in
the same direction or to turn. Therefore, atoms may move chaotically
in the optical lattice. The chaotic transport occurs
if the atomic energy is in the range $0<H<1$. 
At $H<0$, atoms cannot reach even the nearest node and oscillate 
in the first potential well in a regular manner (see Fig.~\ref{fig1}).
At $H>1$, the values of $u$ are always satisfy to the flight condition.
Since the atomic energy is positive in the regime of
chaotic transport, the corresponding conditions can be
summarized as follows: at $|u|<H$,
atom always moves in the same direction, whereas at $|u|>H$,
atom either moves in the same direction, or turns depending on the sign
of $\cos x$ in a given interval of motion. In
particular, if the modulus of $u$ is larger for a long time then the 
energy value, then the atom
oscillates in a potential well crossing two times
each of two neighbor nodes in the cycle.

\begin{figure}[htb]
\begin{center}
\includegraphics[width=0.45\textwidth,clip]{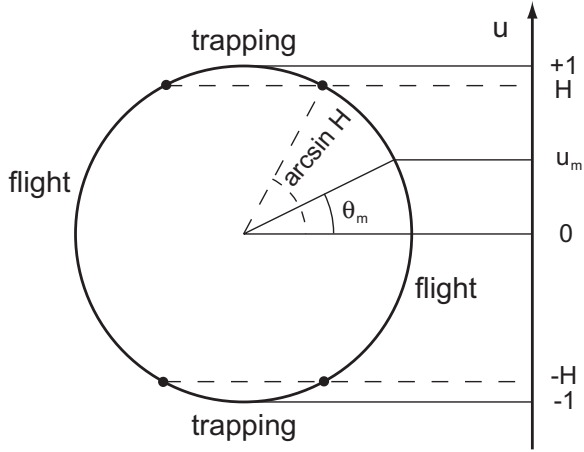}
\end{center}
\caption{Graphic representation for the maps of $u_m$ and
$\theta_m\equiv\arcsin u_m$. $H$ is a given value of the 
atomic energy. Atoms either oscillate in optical potential
wells (trapping) or fly through the optical lattice (flight).
}
\label{circle}
\end{figure}

The conditions stated above allow to find a direct correspondence 
between chaotic atomic transport in the optical
lattice and stochastic dynamics of the Bloch variable $u$.
It follows from Eq. (\ref{u_m}) that the jump magnitude
$u_m-u_{m-1}$ just after crossing the $m$-th node depends
nonlinearly on the previous value $u_{m-1}$.
For analyzing statistical properties of the chaotic
atomic transport, it is more convenient to introduce
the map for $\arcsin u_{m}$
\begin{equation}
\begin{aligned}
\ &\theta_m\equiv\arcsin u_{m}=\\
\ &=\Delta\sqrt\frac{\pi}{\omega_r p_{\rm{node}}}\sin\phi_{m}+\arcsin u_{m-1},
\label{u_ma}
\end{aligned}
\end{equation}
where the jump magnitude does not depend on a
current value of the variable. The map (\ref{u_ma}) visually
looks as a random motion of the point along a
circle of unit radius (Fig.~\ref{circle}). The vertical projection
of this point is $u_m$. The value of the energy $H$ specifies
four regions, two of which correspond  
to atomic oscillations in a well, and two other ones --- 
to ballistic motion in the optical lattice.

We will call ``a flight'' such an event when atom passes, at least,
two successive antinodes (and three nodes). The continuous flight length $L>2\pi$ is
a distance between two successive turning points at which the atom changes
the sign of its velocity, and the discrete flight length is a number
of nodes $l$ the atom crossed. They are related in a simple way,  
$L\simeq\pi l $, for sufficiently long flight.

Center-of-mass oscillations in a well of the
optical potential will be called ``a trapping''. 
At extremely small values of the detuning (the
exact criterion will be given below), the
jump magnitudes are small and the trapping occurs, 
largely, in the $2\pi$-wide wells, i. e., in the space
interval of the length $2\pi$. At intermediate
values of the detuning, it occurs, largely, in the $\pi$-wide
wells, i. e. in the space interval of the length $\pi$.
Far from the resonance, $|\Delta|\gtrsim 1$, trapping
occurs only in the $\pi$-wide wells. Just like to the
case of flights, the number of nodes $l$, atom
crossed being trapped in a well, is a discrete measure
of trapping.
                                                              
\subsection{Statistics of chaotic atomic transport at large jump
magnitudes of $u$}

If the jump magnitudes of the variable $u$ are sufficiently
large
\begin{equation}
|\Delta|\sqrt\frac{\pi}{\omega_r p_{\rm{node}}}\gtrsim\frac{\pi}{2},
\label{largejump}
\end{equation}
then the internal atomic variable $\theta_m\equiv\arcsin u_m$ 
just after crossing the $m$-th node may take with the same probability practically
any value from the range $[-\pi/2,\pi/2]$ (see Fig.~\ref{circle}). With given values
of the recoil frequency $\omega_r=10^{-5}$ and the energy in
the range $0<H<1$ corresponding to chaotically moving atoms, large jumps take place
at medium detunings $|\Delta|\sim 0.1$. The probability
$P_-$ for an atom to turn just after crossing a node
is equal to the probability to get to one of
the trapping regions in Fig.~\ref{circle} (which one depends on 
whether the crossing number $m$ is even or odd) and is given by
\begin{equation}
P_-=\frac{\arccos H}{\pi}<\frac{1}{2} , \quad P_+=1-P_-,
\label{P}
\end{equation}
where $P_+$ is the probability to prolong the motion in the same direction after crossing
the node. It is easily to get from Eq. (\ref{P}) the
probability for an atom to cross $l$ successive
nodes before turning
\begin{equation}
\begin{aligned}
\ &P_{\rm fl}(l )=P_+^{\ l } P_-=\\
\ &=\left(\frac{\arccos H}{\pi}\right)\exp\left[l \ \ln\left(1-\frac{\arccos H}{\pi}\right)\right].
\end{aligned}
\end{equation}
It is a flight probability density function (PDF) in terms of
the discrete flight lengths. The exponential decay means that the atomic transport is normal
for sufficiently large values of the jump magnitudes
of the variable $u$. 

The statistics of the center-of-mass oscillations in the potential wells
can be obtained analogously. With large values of
the jump magnitudes (\ref{largejump}), trapping occurs, largely,
in the $\pi$-wide wells. The probability for a trapped
atom to cross the corresponding well node $l$ times before escaping
from the well is
\begin{equation}
\begin{aligned}
\ &P_{\rm tr}(l )=P_-^{\ l } P_+=\\
\ &=\left(1-\frac{\arccos H}{\pi}\right)\exp\left[l \ \ln\left(\frac{\arccos H}{\pi}\right)\right].
\end{aligned}
\end{equation}

\subsection{Statistics of chaotic atomic transport at small jump magnitudes of $u$}

In this subsection we consider the case of small values of the jump 
magnitudes of the variable $u$ 
\begin{equation}
|\Delta|\sqrt\frac{\pi}{\omega_r p_{\rm{node}}}\ll\frac{\pi}{2}. 
\label{smalljump}
\end{equation}
Now, it may take a long time for an atom to exit from one of the trapping 
or flight regions in Fig.~\ref{circle}. So, we need to calculate the time
of exit of the random variable $\theta_m\equiv\arcsin u_m$ from
one of these regions. The result will depend on what is the length 
of the corresponding circular arc in Fig.~\ref{circle} as compared with 
the jump lengths.

Firstly, let us consider the case if the jump lengths are
small as compared with the lengths both of the flight and trapping
arcs, i. e., if
\begin{equation}
|\Delta|\sqrt\frac{\pi}{\omega_r p_{\rm{node}}}\ll{\rm{min}}\{\arcsin H, 
\arccos H\}.
\label{conds}
\end{equation}
Motion of $\theta_m$ along the circle can be now treated as a one-dimensional
diffusion process for a fictitious particle described by the equation 
\begin{equation}
\frac{\partial{\cal P}(\theta, m)}{\partial m}=D\frac{\partial^2 
{\cal P}(\theta, m)}{\partial \theta^2},
\end{equation}
where ${\cal P}$ is a probability density to find the particle at the angular
position $\theta$ just after crossing the $m$-th node and
$D$ is the corresponding diffusion coefficient of the particle 
\begin{equation}
D\equiv\frac{\aver{(\theta_m-\theta_{m-1})^2}-\aver{\theta_m-
\theta_{m-1}}^2}{2}=\frac{\Delta^2\pi}{4\omega_r p_{\rm{node}}},
\label{diffusion}
\end{equation}
which was calculated for the particle jumping randomly with
the mean magnitude $\aver{\theta_m-\theta_{m-1}}$ (equal to zero in our
case of symmetric jump distribution) and the variance $\aver{(\theta_m-\theta_{m-1})^2}$.
Thus, we reduced the task to the first passage time probability problem for a 
continuous Markov process. Using the results of this theory for a 
Wiener diffusion process \cite{Barucha} described by Eq. 
(\ref{diffusion}), we calculate 
the probability density for a particle to exit from the
interval $\theta_c\pm\theta_{\rm{max}}$ after crossing $l$ nodes 
\begin{equation}
\begin{aligned}
\ &P(l )=\frac{2\pi D}{\theta^2_{\rm{max}}}\sum\limits_{j=0}^{\infty}
(-1)^j(j+1/2)\times\\
\ &\times\cos\left[(j+1/2)\frac{\pi(\theta_0-\theta_c)}{\theta_{\rm{max}}}\right]
\exp\frac{-(j+1/2)^2\pi^2 D l}{\theta^2_{\rm{max}}},
\end{aligned}
\label{33}
\end{equation}
where $\theta_0$ is an initial angular position of the particle in the
region under consideration and $\theta_c$ is the center of the region
(with four possible values $0$, $\pi/2$, $\pi$, $3\pi/2$). 

Since in a case of small jumps a particle gets
to the region near its limit point,
it is possible to replace $\theta_0$ by the quantity 
$\theta_0=\theta_c-\theta_{\rm{max}}+\epsilon$ with
a small positive value of $\epsilon$. Expanding the cosine in (\ref{33}) 
in the vicinity of the nodes
and taking into account only the terms of the first order of smallness, we get
the flight and trapping PDFs
\begin{equation}
\begin{aligned}
P_{\rm fl}(l )\simeq\frac{Q}{\arcsin^3 H}&\sum\limits_{j=0}^{\infty}(j+1/2)^2\times\\
&\times\exp\frac{-(j+1/2)^2\pi^2 D l }{\arcsin^2 H},\\
P_{\rm tr}(l )\simeq\frac{Q}{\arccos^3 H}&\sum\limits_{j=0}^{\infty}(j+1/2)^2\times\\
&\times\exp\frac{-(j+1/2)^2\pi^2 D l }{\arccos^2 H},
\label{stat}
\end{aligned}
\end{equation}
where $Q$ is a normalization constant that is the
same in both the cases. When deriving the first and second 
expressions (\ref{stat}), we replaced 
 $\theta_{\rm{max}}$ by $\arcsin H$ and $\arccos H$, respectively. 
It is easily to realize
that when $j$ exceeds the value $\theta_{\rm{max}}/(\pi\sqrt{Dl})$, the
corresponding terms of the series (\ref{stat})
decrease rapidly.

The PDFs (\ref{stat}) can be written in a much
more simple form for two conditions imposed on the
number of jumps (node crossings) $l$ a particle needs to quit
the interval $\theta_c\pm\theta_{\rm{max}}$. If $l\gtrsim 
\theta^2_{\rm{max}}/D$, then
all the terms in the sums (\ref{stat}) are small as compared
with the first one. Both the flight
and trapping statistics are exponential in this case.

To the contrary, if $l\ll \theta^2_{\rm{max}}/D$, then one should
take into account a large number of terms in the
sums (\ref{stat}), and each sum can be replaced
approximately by the integral. In this case the result does not depend on 
the length of the region $2\theta_{\rm{max}}$ and we get 
the power law decay 
\begin{equation}
\begin{aligned}
P(l )\simeq\frac{Q}{\theta^3_{\rm{max}}}\int\limits_{0}^{\infty}(j+1/2)^2
\exp\frac{-(j+1/2)^2\pi^2 D l }{\theta^2_{\rm{max}}}dj\simeq\\
\simeq\frac{Q\pi^{-2.5}D^{-1.5}}{4} l ^{-1.5},\quad l\ll \frac{\theta^2_
{\rm{max}}}{D}
\label{38}
\end{aligned}
\end{equation}
both for the flight and trapping PDFs. The power-law statistics (\ref{38})
implies anomalous atomic transport. Dividing the
length $2\theta_{\rm{max}}$ by the jump magnitude (\ref{smalljump}),
we obtain the quantity 
\begin{equation}
l_{\rm{cr}}\equiv\theta_{\rm{max}}/\sqrt{D}
\end{equation}
that is a minimum
number of jumps (node crossings) a particle needs to pass the
region through. With the number of jumps $l <l_{\rm{cr}}$,
a particle, randomly moving on the circle, may get 
out of the interval $\theta_c\pm\theta_{\rm{max}}$ only through the
same border where it got in. The statistics of exit times is
known to be a power-law one with the transport
exponent equal to $-1.5$. The length of the interval
does not matter in this case. If $l \gtrsim l_{\rm{cr}}$, then
particles may exit through both the borders, and the
corresponding statistics cannot be approximated by a single 
transport exponent. If $l \gtrsim l^2_{\rm{cr}}\equiv\theta^2_{\rm{max}}/D$, 
then we expect an exponential decay.

The size of the trapping and flight regions
depends on the value of the atomic energy $H$ (see Fig.~\ref{circle}).
At $H>\sqrt{2}/2$ ($\arcsin H>\pi/4$), the flight PDF has
a longer decay than the trapping PDF. On the contrary,
at $H<\sqrt{2}/2$, the $P_{\rm tr}$'s decay is longer than the $P_{\rm fl}$'s one.
If the jump magnitude (\ref{smalljump}) is of the order of the
size of the flight or trapping regions
\begin{equation}
\begin{aligned}
|\Delta|\sqrt\frac{\pi}{\omega_r p_{\rm{node}}}\sim\arcsin H\ll\frac{\pi}{2},\\
{\rm{or}}\quad\quad\\
|\Delta|\sqrt\frac{\pi}{\omega_r p_{\rm{node}}}\sim\arccos H\ll\frac{\pi}{2},
\end{aligned}
\label{rapidpass}
\end{equation}
then a particle may pass through the region making a
small number of jumps $l$. So, the approximation of the
diffusion process (\ref{conds}) fails, and the corresponding 
PDF is exponential.

\begin{figure}[htb]
\begin{center}
\includegraphics[width=0.45\textwidth,clip]{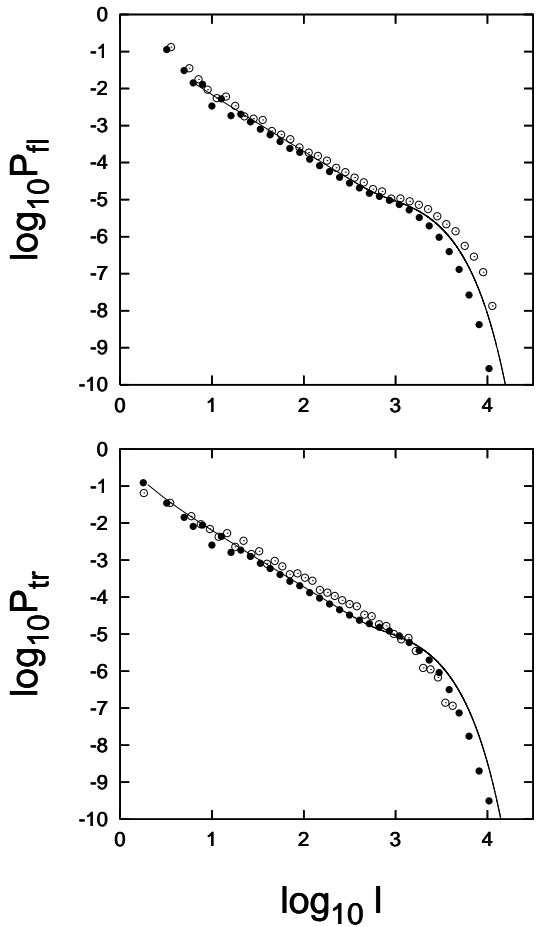}
\end{center}
\caption{The flight  $P_{\rm fl}$ and trapping $P_{\rm tr}$ PDFs for a chaotically
moving atom with comparatively small values of the jump magnitudes 
of the variable $u$. The detuning $\Delta=-0.001$ is small, and the energy value 
$H=0.724$ ($p_0=535$) provides approximately equal sizes of the flight and trapping 
regions in Fig.~\ref{circle}. White and black circles represent results
of integration of the basic (\ref{mainsys}) and reduced (\ref{redsys})
equations of motion, respectively, and the solid lines 
represent the analytical PDFs (\ref{stat}). 
$x_0=0$, $z_0=-1$, $u_0=v_0=0$, $\omega_r=10^{-5}$.}
\label{stats1}
\end{figure}

In order to check the analytical results obtained in this section, we 
compare them with numerical simulation of the reduced
(\ref{redsys}) and basic (\ref{mainsys}) equations of motion. 
The reduced equations (\ref{redsys}) describe the center-of-mass motion 
modulated by a jump-like behavior  of the internal variable $u$ which 
satisfies to the stochastic map (\ref{u_m}).  
To explore different regimes of the chaotic atomic transport we  
integrate Eqs. (\ref{redsys}) and (\ref{mainsys}) numerically with two
values of the detuning $\Delta=-0.001$ and $\Delta=-0.01$ (the
cases of small and medium jump magnitudes of
the variable $u$, respectively) and different
values of the atomic energy and momentum and 
compute the PDFs of the flight and
trapping events. They are normalized histograms
for a number of standing-wave nodes $l$ which the  atom crossed in a run.
Each PDF is computed with a single but very long (up to $\tau \sim 10^8$ 
for the basic equations of motion) atomic trajectory.  

\begin{figure}[htb]
\begin{center}
\includegraphics[width=0.45\textwidth,clip]{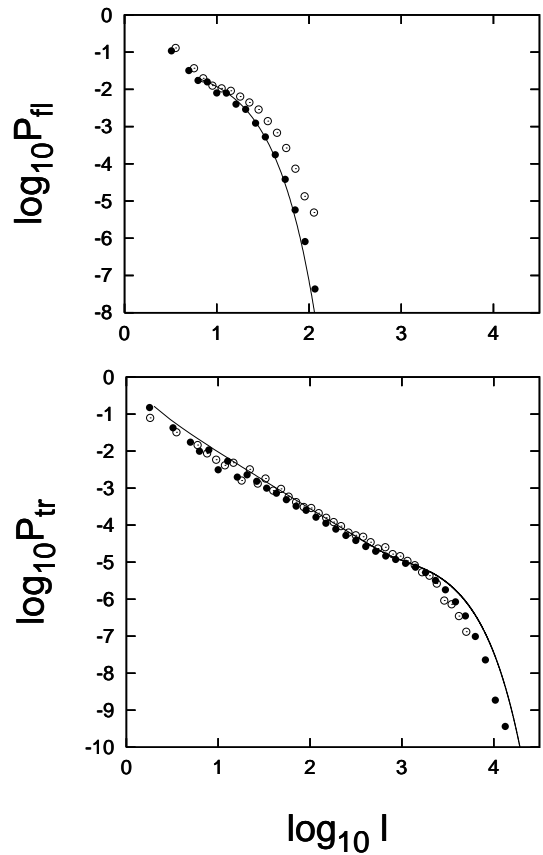}
\end{center}
\caption{The same as in Fig.~\ref{stats1} but with the energy 
$H=0.101$ ($p_0=402$) providing a narrow flight  region as compared with 
the trapping one.}
\label{stats2}
\end{figure}

In Fig.~\ref{stats1} we compare the results (in a log-log scale) in the case of
small jump magnitudes of the variable $u$ ($\Delta=-0.001$)
and approximately equal sizes of the flight and
trapping regions ($H=0.724\sim\sqrt{2}/2$). White
and black circles represent the results
of numerical integration of the basic
(\ref{mainsys}) and reduced (\ref{redsys}) equations
of motion, respectively, and the solid line
represents the analytical predictions (\ref{stat}). The agreement between 
the PDFs computed with Eqs. (\ref{mainsys}), (\ref{redsys}), and (\ref{stat}) is 
rather good. All the flight and trapping PDFs exhibit in their dependence on 
the crossing number $l$ three kinds of decay which are fairly 
approximated by the formulas (\ref{stat}). The critical value of the 
crossing number $l$ is 
estimated to be $l_{\rm{cr}}\simeq 55$ with chosen values of the parameters 
and initial conditions.  In the range $l \lesssim 55$ the PDFs 
are expected to demonstrate the power law decay with 
the slope $-1.5$ given by the formula (\ref{38}). In the range 
$55 \lesssim l \lesssim l^2_{\rm{cr}} \simeq 3000$, there should exist a number 
of different transport exponents. The initial part of this range
is fitted by the same power-law function. However, the other part of the
range cannot be fitted by a simple function. In the range 
$l \gtrsim l^2_{\rm{cr}} \simeq 3000$, the decay is expected to be 
purely exponential in accordance with the 
the first expression in Eqs. (\ref{stat}).   

In Fig.~\ref{stats2} we show the PDFs computed at the same detuning 
$\Delta=-0.001$ as in Fig.~\ref{stats1} but with the value of the atomic
energy $H=0.101$ for which the flight region in Fig.~\ref{circle}  
is narrower than the trapping one. The jump magnitude is now of the order 
of the length of the flight region (see (\ref{rapidpass})) and the 
flight PDFs are expected to be exponential ($l_{cr}\simeq 9$). The critical value of the 
crossing number $l$ for trapping is $l_{\rm{cr}}\simeq 58$ 
with given values of the energy and the detuning. Therefore, 
in the range $l \lesssim 58$ the trapping PDFs 
are expected to vary as the power law decay with 
the slope $-1.5$ given by the formula (\ref{38}). In the range 
$l \gtrsim l^2_{\rm{cr}} \simeq 3400$, the decay is expected to be 
purely exponential in accordance with the the second expression 
in Eqs. (\ref{stat}). The trapping PDFs in the range 
$58 \lesssim l \lesssim l^2_{\rm{cr}} \simeq 3400$
are neither power law nor exponential demonstrating a complicated behavior. 
This speculation is confirmed by the numerical computation shown 
in Fig.~\ref{stats2}.  
 
\begin{figure}[htb]
\begin{center}
\includegraphics[width=0.45\textwidth,clip]{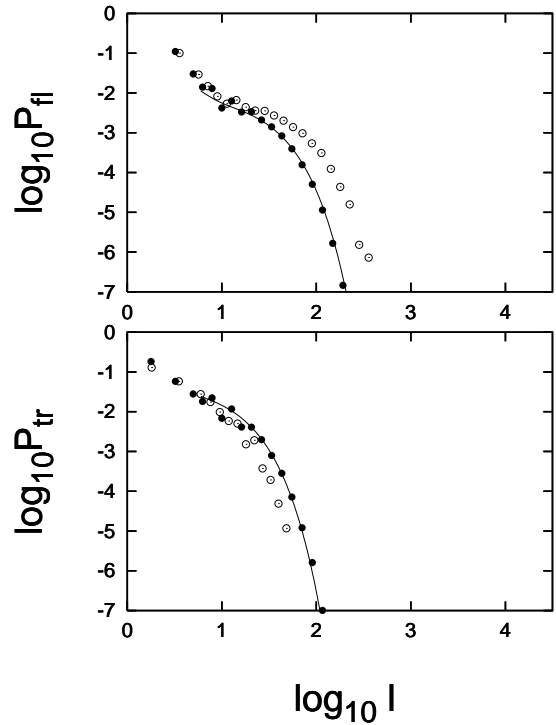}
\end{center}
\caption{The flight  $P_{\rm fl}$ and trapping $P_{\rm tr}$ PDFs for a chaotically
moving atom with comparatively large values of the jump magnitudes 
of the variable $u$. The detuning $\Delta=-0.01$ is medium, and the energy value 
$H=0.8055$ ($p_0=550$) provides a domination of the flight events over the 
trapping ones. White and black circles represent results
of integration of the basic (\ref{mainsys}) and reduced (\ref{redsys})
equations of motion, respectively, and the solid lines 
represent the analytical PDFs (\ref{stat}). 
$x_0=0$, $z_0=-1$, $u_0=v_0=0$, $\omega_r=10^{-5}$.}
\label{stats3}
\end{figure}
In order to demonstrate what happens with larger
values of the jump magnitudes, we take the detuning to be $\Delta=-0.01$  
increasing the jump magnitude in ten times as compared with the 
preceding cases. With the taken value of the energy
$H=0.8055$ we provide a slight domination of flights over trappings. 
The jump magnitude is now so large that particles may pass both through 
the flight and trapping regions making a small number of jumps. 
It is expected that all the PDFs, both the flight and trapping ones, 
should be practically exponential in the whole range of the crossing number $l$. 
It is really the case (see Fig.~\ref{stats3}).

\section{Dynamical fractals}

Various fractal-like structures may arise in chaotic
Hamiltonian systems \cite{Ott, 13, 14}. In our previous papers
\cite{JETP, JRLR} we have found numerically fractal properties
of chaotic atomic transport in cavities and optical lattices.
In this section we apply the analytical results of the theory
of chaotic transport, developed in the preceding
sections, to find the conditions
under which the dynamical fractals may arise.

We place atoms one by one at the point $ x_0=0$
with a fixed positive value of the momentum $p_0$ and compute the time $T$ when they cross one
of the nodes at $x=-\pi/2$ or $x=3\pi/2$. In these
numerical experiments we change the value of the
atom-field detuning $\Delta$ only. All the 
initial conditions $p_0=200$, $z_0=-1$, $u_0=v_0=0$ and
the recoil frequency $\omega_r=10^{-5}$ are fixed. 
\begin{figure}
\begin{center}
\includegraphics[width=0.48\textwidth,clip]{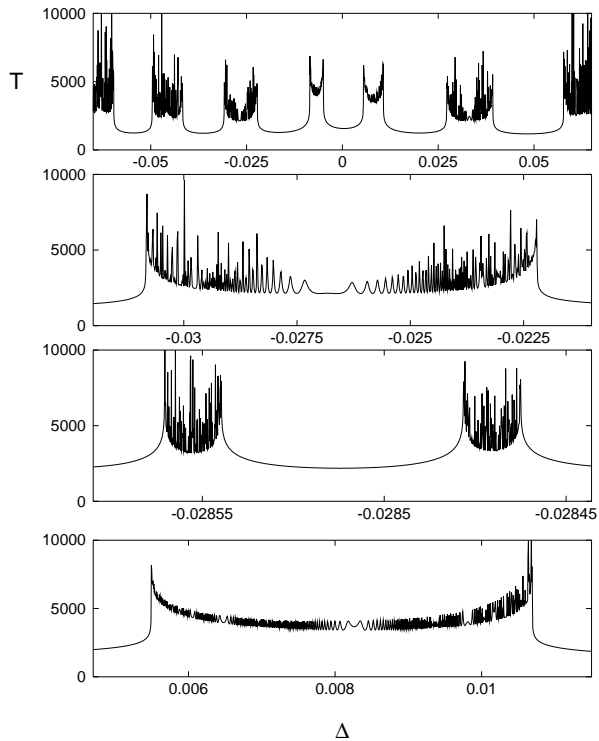}
\end{center}
\caption{Fractal-like dependence of the time of exit of atoms $T$
from a small region in the optical lattice on the detuning $\Delta$:
$p_0=200$, $z_0=-1$, $u_0=v_0=0$. Magnifications of the detuning 
intervals are shown.}
\label{fig8}
\end{figure}
The exit time function $T(\Delta)$ in Fig.~\ref{fig8} demonstrates an intermittency of smooth
curves and complicated structures that cannot be resolved in principle, no
matter how large the magnification factor. The second and third panels in
Fig.~\ref{fig8}
demonstrate successive magnifications of the detuning intervals shown
 in the upper panel.
Further magnifications reveal a self-similar fractal-like structure that
is typical for Hamiltonian systems with chaotic scattering \cite{Ott, 14}.
The exit time
$T$, corresponding to both smooth and unresolved $\Delta$ intervals, increases
with increasing the magnification factor. Theoretically, there exist atoms
never crossing the border nodes at $x=-\pi/2$ or $x=3\pi/2$ in spite of the 
fact that they have no obvious
energy restrictions to do that. Tiny interplay between chaotic external
and internal atomic dynamics prevents those atoms from leaving the small space. 
The similar
phenomenon in chaotic scattering is known as dynamical trapping.
In Ref. \cite{JETP} we have computed the
Hausdorff dimension for the similar fractal and shown that it is not an integer.
We note that the fractal-like structures similar to that shown in 
Fig.~\ref{fig8} arise in numerical experiments with longer distances 
between the border nodes but the corresponding computation time increases 
significantly with increasing this distance. 

Various kinds of atomic trajectories can be
characterized by the number of times $m$ atom crosses the central node at 
$x=\pi/2$ between the border nodes.
There are also special separatrix-like
trajectories along which atoms asymptotically reach the points
with the maximum of the potential energy, having no more kinetic energy to
overcome it. In difference from the separatrix motion in the resonant 
system ($\Delta=0$)
with the initial atomic momentum $p_{\rm {cr}}$, a detuned atom can
asymptotically reach one of the stationary points even if it was trapped for a while
in a well. We specify the $mS$-trajectory as a trajectory
of an atom crossing the central node $m$ times before starting
the separatrix-like motion. Such an asymptotical motion
takes an infinite time, so the atom will never reach the border nodes.

The smooth $\Delta$ intervals in the first-order
structure (Fig.~\ref{fig8}, upper panel) correspond to atoms which never
change the direction of motion ($m=1$) and reach the border node at $x=3\pi/2$.
The singular
points in the first-order structure with $T=\infty$, which are located at the border between the
smooth and unresolved $\Delta$ intervals, are generated by the
$1S$-trajectories. Analogously, the smooth and unresolved $\Delta$ intervals
in the second-order structure (second panel in Fig.~\ref{fig8}) correspond to
the 2-nd order ($m=2$) and the other trajectories, respectively, with singular points between them
corresponding to the $2S$-trajectories, and so on.

There are two different mechanisms of generation of infinite exit times,
namely,
dynamical trapping with an infinite number of oscillations ($m=\infty$) 
inside the interval $[-\pi/2, 3\pi/2]$ and the
separatrix-like motion ($m\ne\infty$). The set of all the values of the detunings, generating
the separatrix-like trajectories, was shown to be a countable fractal in 
Refs. \cite{JETP, JRLR},
whereas the set of the values generating dynamically trapped atoms with
$m=\infty$ seems to be uncountable. The exit time 
$T$ depends  in a complicated way not only on the values of the 
control parameters but on 
initial conditions as well. In Ref. \cite{JRLR} we presented a two-dimensional image
of this function in two coordinates of
the initial atomic momentum $p_0$ and the atom-field detuning $\Delta$.

The fractal-like structure with smooth and unresolved components
may appear if atoms have an alternative either to turn back or to prolong the 
motion in
the same direction just after crossing the node
at $x=\pi/2$. For the first-order structure in
the upper panel in Fig.~\ref{fig8}, it means that the internal
variable $u$ of an atom, just after crossing the node
for the first time ($\cos x<0$), satisfies either to the condition $u_1<H$
(atom moves in the same direction), or to the condition $u_1>H$
(atom turns back). If $u_1=H$, then the exit time $T$ is
infinite. The jumps of the variable $u$ after crossing the node
are deterministic but sensitively dependent
on the values of the control parameters and initial
conditions. We have used this fact when introducing the stochastic map 
in \setcounter{abcd}{4}Sec.~\Roman{abcd}. 
Small variations in these values lead to
oscillations of the quantity $\arcsin u_1$ around the initial value $\arcsin u_0$ with the
amplitude $|\Delta|\sqrt{\pi/(\omega_r p_{\rm{node}})}$. If this amplitude is large
enough, then the sign of the quantity $u_1-H$ alternates and we obtain alternating
smooth (atoms reach the border $x=3\pi/2$ without changing their direction of
motion) and unresolved (atoms turns a number of times
before exit) components of the fractal-like structure.
The singular points at the border of such components
correspond to the atoms with infinite exit time.

If the values of the parameters admit large jump
magnitudes of the variable $u$ (see condition (\ref{largejump})), then
the dynamical fractal arises in the energy range $0<H<1$, i. e., at the
same condition under which atoms move in the optical lattice 
in a chaotic way. In a case of small jump magnitudes,
fractals may arise if the initial value of an atom $u_0$ is close 
enough to the value of the energy $H$, i.e., the atom has 
a possibility to overcome the value $u=H$ in one jump. Therefore,
the condition for appearing in the fractal $T(\Delta)$ the first-order 
structure with singularities is the following:
\begin{equation}
|\arcsin u_0-\arcsin H|<|\Delta|\sqrt{\frac{\pi}{\omega_r p_{\rm{node}}}}.
\label{cond1}
\end{equation}

The formation of the second-order structure is explained analogously. 
If an atom made a turn
after crossing the node for the first time, then it will cross the node for 
the second time. After that, the atom either will turn or cross the border 
node at $x=-\pi/2$. What will happen depend on the value of $u_2$. 
However, in difference from the case with $m=1$, the condition for appearing  
an infinite exit time with $m=2$ is $u_2=-H$.
Furthermore, the previous value $u_1$ is not fixed  (in difference from $u_0$) 
but depends on the value of the detuning $\Delta$. In any case we have 
$u_1>H$ since the second-order structure consists of the trajectories of those 
atoms which turned after the first node crossing. Near the singular borders
of the second-order structure we have $u_1\simeq H$, whereas far from them 
$u_1$ may be much larger. In order for an atom would be able to turn 
after the second node crossing, the magnitude of its variable 
$u$ should change sufficiently to be in the range $u_2<-H$. The atoms, whose  
variables $u$ could not ``jump'' so far, leave the space $[-\pi/2, 3\pi/2]$. 
The singularities are absent in the middle part of 
the second-order structure shown in the second panel in Fig.~\ref{fig8}
because all the corresponding atoms left the space after the second node crossing. 
The variable $u_2$ oscillates with varying $\Delta$ generating oscillations 
of the exit time. In vicinity of the borders of the second-order structure   
the required jump magnitude $u_2-u_1$ is smaller and is equal approximately to 
$2H$. Some atoms could turn for the second time or follow  
separatrix-like trajectories which generate singularities of the exit-time 
function. So, the condition for appearing singularities in 
the second-order structure is the following:
\begin{equation}
2 \arcsin H<|\Delta|\sqrt{\frac{\pi}{\omega_r p_{\rm{node}}}}.
\label{cond2}
\end{equation}
With the values of the parameters taken in the simulation, 
we get the energy $H=0.2+\Delta/2$. It is easy
to obtain from the inequality (\ref{cond2}) the approximate value of the 
detuning $|\Delta|\approx0.0107$ for which the second-order
singularities may appear. In the lower panel in Fig.~\ref{fig8}
one can see this effect. No additional conditions are required for 
appearing the structures of the third and the next orders. 

It should be noted that the inequality (\ref{cond2})
is opposite to the first part of the inequality (\ref{conds})
that determines the condition for appearing power law decays in the flight
PDF. Therefore, dynamical fractal may appear in those ranges of
the control parameters where the L\'evy flights are impossible and vice versa.
However, the trapping PDF may have a power law decay. It should
be noted that the inequality (\ref{cond2}) in difference from (\ref{cond1}) 
is strongly
related with the chosen concrete scheme for computing exit times.
It is not required with other schemes, say, with three antinodes 
between the border nodes. 

\section{Conclusion and discussion}

We have studied semiclassical transport of atoms in a one-dimensional
optical lattice in the framework of the idealized 
model of the atom-field interaction. The center-of-mass
motion is strongly affected by the evolution of the atomic
Bloch variables --- the synchronized and quadrature components of the atomic 
dipole moment
$u$ and $v$ and the population inversion $z$. There are
ranges of the atom-field detuning $\Delta$ and the initial
momentum $p_0$ where the atomic transport has been shown
to be chaotic with a complicated alternation of
trappings in wells of the optical potential and
flights over them. We developed a semiclassical theory of
the chaotic atomic transport in optical lattices in terms of a random walk 
of the Bloch variable $u$ and
confirmed the analytical results by direct numerical simulation of 
the basic and reduced equations of motion. 

Based on a jump-like behavior of the internal
variable $u$ for atoms crossing nodes of the standing
laser wave, we constructed a stochastic map that
determines the center-of-mass motion. We found the
relations between the detuning $\Delta$, recoil frequency $\omega_r$, and
the atomic energy $H$, under which atoms may move in the lattice
in a chaotic way. To illustrate statistical properties of the chaotic transport,  
we proposed a simple model of random walking of the quantity $\arcsin u$
on a circle.

The statistical properties of chaotic atomic
transport strongly depend on the relation between the
jump magnitude of the variable $u$ and the atomic energy.
Both the length of flight and the duration of trapping 
are measured by a number of nodes $l$ atom crossed
during the flight and trapping events, respectively. 
The PDFs for the flight $P_{\rm fl}(l)$ and
trapping $P_{\rm tr}(l)$ events were analytically derived to be exponential in
a case of large jumps. In a case of small jumps, 
the kind of the statistics depends on additional conditions imposed 
on the atomic and lattice parameters, and the distributions 
$P_{\rm fl}(l)$ and $P_{\rm tr}(l)$ 
were analytically shown to be either practically 
exponential or functions with 
long power law parts with the slope $-1.5$ but exponential ``tails''.
The comparison of the PDFs computed
with analytical formulas, the stochastic map, and
the basic equations of motion has shown a good agreement in different ranges 
of the atomic and lattice parameters.
We used the results obtained to find the analytical conditions, 
under which the
fractal properties of the chaotic atomic transport were observed, and
to explain the structure of the corresponding dynamical fractals.

Since the period and amplitude of the optical potential and the atom-field 
detuning can be modified in a controlled way, the transport exponents of 
the flight and trapping distributions are not fixed but can be varied 
continuously, allowing to explore different regimes of the atomic transport. 
Our analytical and numerical results with the idealized system have shown 
that deterministic atomic transport in an optical lattice cannot be just 
classified as normal and anomalous one. We have found that the flight and
trapping PDFs may have long algebraically decaying parts and a short 
exponential ``tail''. It means that in some ranges of the atomic and lattice 
parameters numerical experiments reveal  
anomalous transport with L\'evy flights (see Figs.~\ref{stats1} 
and ~\ref{stats2}). The 
transport exponent equal to $-1.5$ means that the first, second, and the other 
statistical moments are infinite for a reasonably long time. The corresponding atomic trajectories 
computed for this time are self-similar and fractal. The total distance, 
that the atom travels for the time when the flight PDF
 decays algebraically, is dominated by a single flight. However, 
the asymptotical behavior is close to normal transport. 
In other ranges of the atomic and lattice parameters, the transport is 
practically normal both for short and long times (see Fig.~\ref{stats3}). 

In conclusion we should emphasize that the theory of chaotic atomic transport 
in an optical lattice presented in this paper does not take into account
spontaneous emission which is inevitable in any real experiment. Speculation 
about small changes in the atomic momentum of the order of $\hbar k_f$ 
(caused by spontaneous emission) as compared with its initial values of the order of 
$300 \div 500 \hbar k_f$ used in our computations is not a convincing 
argument because the atomic momentum is equal to zero at turning points, and  
spontaneous emission may change the center-of-mass motion in vicinity of these 
points. Spontaneous emission can be incorporated in the equations of motion 
of the type of our basic equations (\ref{mainsys}) \cite{AP06} 
by the quantum Monte Carlo method \cite{Carmichael}. In general, we expect 
a competition between the coherent but chaotic atomic dynamics and 
random processes of spontaneous emission. In particular, the stochastic map 
(\ref{u_m}) constructed in this paper should be generalized to include jumps 
of the variable $u$ to the zero value  at random time moments. What will 
happen with the flight and trapping PDFs is an open question. We plan to study 
the atomic transport in an optical lattice with spontaneous emission 
in the future.

\section{Acknowledgments}
This work was supported  by the Russian Foundation for Basic Research
(project no. 06-02-16421), by the Program "Mathematical methods in
nonlinear dynamics"of the Prezidium of the Russian Academy of Sciences, 
and the program of the Prezidium
of the Far-Eastern Division of the Russian Academy of Sciences.

\end{document}